\documentclass[12pt]{article}
\usepackage[psamsfonts]{euscript}
\usepackage[psamsfonts]{amssymb}
\usepackage{amsmath}
\usepackage{graphicx}
\usepackage[small,sc,center]{titlesec}
\usepackage{indentfirst}

\newcommand{\nc}{\newcommand}
\nc{\EK}   {E_\mathrm{K}}
\nc{\K}    {\:\mathrm{K}}
\nc{\dmbol}{\Delta M_\mathrm{bol}}
\nc{\mzams}{M_\mathrm{ZAMS}}
\nc{\Teff} {T_\mathrm{eff}}
\nc{\xs}   {X_\mathrm{s}}
\nc{\ys}   {Y_\mathrm{s}}
\begin{document}

\begin{center}
\textbf{INSTABILITY OF LBV--STARS AGAINST RADIAL OSCILLATIONS}

\vskip 3mm
%\copyright\quad
\textbf{2009 \quad Yu. A. Fadeyev\footnote{e--mail: fadeyev@inasan.ru}}

\textit{Institute of Astronomy, Moscow} \\

% Submitted: May, 2009
\end{center}

\vskip 10pt\noindent
In this study we consider the nonlinear radial oscillations
exciting in LBV--stars with effective temperatures
$1.5\cdot 10^4\K\le\Teff\le 3\cdot 10^4\K$,
bolometric luminosities
$1.2\cdot 10^6L_\odot\le L\le 1.9\cdot 10^6L_\odot$
and masses $35.7 M_\odot\le M\le 49.1 M_\odot$.
Hydrodynamic computations were carried out with initial conditions
obtained from evolution sequences of population~I stars
($X=0.7$, $Z=0.02$) with initial masses in the range
$70M_\odot\le\mzams\le 90M_\odot$.
All models show instability against radial oscillations
with amplitude growth time comparable with dynamical time scale of the
star.
Radial oscillations exist in the form of nonlinear running waves
propagating from the boundary of the compact core to the
upper boundary of the hydrodynamical model.
The velocity amplitude of outer layers is of several hundreds of km/s
while the bolometric light amplitude is $\Delta M_\mathrm{bol}\le 0.2$~mag.
Stellar oscillations are not driven by the $\kappa$--mechanism and
are due to the instability of the gas with adiabatic exponent
close to the critical value $\Gamma_1 = 4/3$ due to the large
contribution of radiation in the total pressure.
The range of light variation periods
($6~\mathrm{day}\le\Pi\le 31~\mathrm{day}$) of hydrodynamical models
agrees with periods of microvariability observed in LBV--stars.



% \newpage
\section*{introduction}

The most luminous ($L \sim 10^6 L_\odot$) stars represent a group of
luminous blue variables (LBV) which involve such well known objects
as $\eta$~Car, P~Cyg, S~Dor and the Hubble--Sandage variables in
galaxies M31 and M33.
All LBV--stars are massive population~I stars at the early helium
burning stage and their small number is due to the rapid evolutionary
movement across the HRD to the higher effective temperatures.
LBV--stars are thought to precede the Wolf--Rayet evolutionary stage
(Maeder, 1983; Langer et al., 1994) though it is not excepted that
they might also be the supernova progenitors
(Gal--Yam et al., 2007; Trundle et al., 2008).

Photometric variability of LBV--stars can be divided into three distinct
types depending on the cha\-racteristic time scale
(Lamers, 1987; Leitherer et al., 1992).
In variability of the first type (giant eruptions)
the bolometric luminosity increases by several magnitudes and
the mass of the ejected material is as high as $\sim M_\odot$
(Humphreys \& Davidson, 1994).
The energy of the giant eruption can be so high that the LBV--star
becomes the SN--impostor (Smith et al., 2009).
There are known two giant eruption events in the Galaxy.
One of them was P~Cyg in the early XVII century (de Groot, 1988)
and another was $\eta$~Car in the middle of XIX century
(van Genderen, 1984; Frew, 2004),
so that the time scale of giant eruptions is thought to be of
the order of $\gtrsim 10^2$~yr.
In variability of the second type (S~Dor type) with time scale of
$\sim 10$~yr the bolometric light seems do not change, whereas the
visual light changes by about 2~mag (van Genderen, 2001).
During the absence of S~Dor type variability one can observe
the cyclic light variations (microvariability) with amplitude
of $\le 0.2$~mag
and characteristic time ranging from one to several dozens of day.

The origin of the both giant eruptions and S~Dor variability
remains still unclear and only microvariations are interpreted in terms
of stellar pulsations (van Genderen, 1989).
This assumption is supported by the linear theory of stellar pulsation
(Glatzel \& Kiriakidis, 1993;
Kiriakidis et al., 1993; Dziembowski \& Slawinska, 2005)
which predicts that outer layers of LBV--stars are unstable against
radial oscillations.
This fact seems to be of great importance because observational estimates of
microvarion periods can provide us with independent method of mass determination.

The goal of the present study is to consider nonlinear radial stellar
oscillations as a cause of microvariability in LBV--stars.
Hydrodynamic modeling of this phenomenon is too complicated
because of the large ampliture oscillations exciting at the boundary
of the dynamical instability, so that only two reports on this item
have been published by now (Dorfi et al., 2000; Dorfi \& Gautschy, 2002).
Below we present the results of hydrodynamic calculations of radial
oscillations in LBV--stars with effective temperatures
$1.5\cdot 10^4\K\le\Teff\le 3\cdot 10^4\K$ and luminosities
$1.2\cdot 10^6 L_\odot\le L \le 1.9\cdot 10^6 L_\odot$.
Initial conditions for the Cauchy problem of the equations of
radiation hydrodynamics were taken from the stellar evolution
calculations for population~I stars ($X=0.7$, $Z=0.02$)
with initial masses $70M_\odot\le\mzams\le 90M_\odot$.
This paper continues our earlier studies of nonlinear radial
oscillations in the massive helium--burning stars
(Fadeyev, 2007; 2008a; 2008b)
where the computational methods are described in more detail.

\section*{evolutionary models}

In the present study the stellar evolution during the hydrogen core burning
was calculated with mass loss rates $\dot M$ by Vink et al. (2000, 2001)
that are based on the stellar wind models with multiple photon scattering and that
are in a good agreement with observations of O and B stars.
Comparison with our previous calculations (Fadeyev, 2007, 2008a, 2008b) shows
that application of the formula by Nieuwenhuijzen and de Jager (1990)
for the hydrogen burning phase leads to much higher (by a factor of $\lesssim 20$)
mass loss rates and, therefore, to substantially smaller stellar mass and
luminosity at the core hydrogen exhaustion.
For example, evolution calculation for the star $\mzams=80M_\odot$
up to the central hydrogen abundance $X_c = 10^{-5}$ gives
the stellar mass and luminosity $M=43.6M_\odot$ and $L=1.27\cdot 10^6L_\odot$
with mass loss rates by Nieuwenhuijzen and de Jager (1990),
whereas from calculations with mass loss rates by Vink et al. (2000, 2001)
the mass and luminosity are $M=50.9M_\odot$ and $L=1.55\cdot 10^6L_\odot$.

The formulae for $\dot M$ by Vink et al. (2000, 2001) become
inapplicable after the core hydrogen exhaustion and following
Vazquez et al., (2007) the mass loss rates at later evolutionary
stages were calculated according to Nieuwenhuijzen and de Jager (1990).
During the helium core burning when the effective temperature
rises above $\Teff = 10^4\K$ the mass loss rates were calculated
according to Nugis and Lamers (2000).

\begin{figure}
\centerline{\includegraphics[width=16cm]{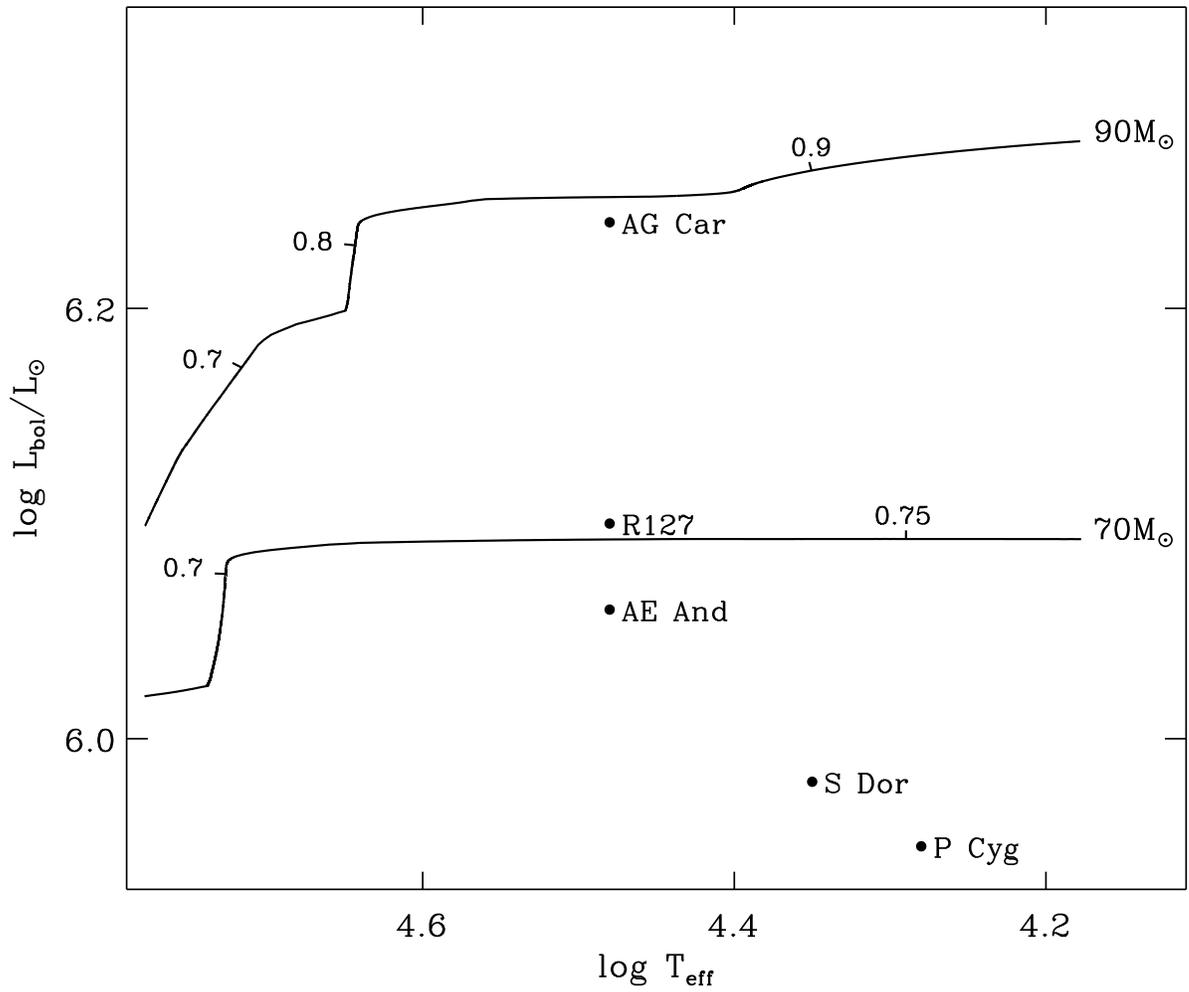}}
\caption{LBV--stars on the HRD (according to de Jager (1998)) and
parts of the evolutionary tracks
$\mzams=70M_\odot$ and $\mzams=90M_\odot$.
Attached at the curves are the values of the central helium abundance $Y_c$.
}
\label{fig1}
\end{figure}

Fig.~\ref{fig1} displays the HRD with several LBV--stars
with luminositiues and effective temperatures from de Jager (1998).
On the same figure are shown the parts of evolutionary tracks
with initial masses $\mzams=70M_\odot$ and $\mzams=90M_\odot$.
The tracks cross the HR--diagram from the right to the left
and labels at the tracks give the central helium mass fraction $Y_c$.
Within the displayed parts of the tracks the central temperature $T_c$ and
the central gas density $\rho_c$ change negligibly.
For all models $T_\mathrm{c}\approx 2.15\cdot 10^8\K$ while
the central gas density ranges within
$\rho_\mathrm{c}\approx 250~\mathrm{cm}^3/\mathrm{g}$ for $\mzams=90M_\odot$
and $\rho_\mathrm{c}\approx 300~\mathrm{cm}^3/\mathrm{g}$  for $\mzams=70M_\odot$.

For $\mzams=70M_\odot$
the time needed to cross the effective temperature range
$1.5\cdot 10^4\K\le\Teff\le 5\cdot 10^4\K$
is $\approx 8\cdot 10^3$~yr and the stellar
mass decreases from $36M_\odot$ to $31M_\odot$.
For $\mzams=90M_\odot$ the evolution time within the
same effective temperature range is $\approx 5\cdot 10^4$~yr
and the stellar mass decreases from $49M_\odot$ to $36M_\odot$.
Vink and de Koter (2003) evaluated the mass of AG~Car as
$M\approx 35M_\odot$.
Bearing in mind existing uncertainties in estimates of the
luminosities of LBV--stars one may conclude that evolutionary
sequences computed in the present study are in a good agreement
with observations.

\section*{hydrodynamic models}

Some evolutionary stellar models corresponding to the early helum
core burning were used as initial conditions in hydrodynamic
computations of the self--exciting radial stellar oscillations.
Main parameters of the hydrostatically equilibrium models
are given in the table, where $R$ is the equilibrium radius of photosphere.
For effective temperatures
$1.5\cdot 10^4\K\le\Teff\le 3\cdot 10^4\K$
the mass of the enevelope surrounding the compact core is
almost negligible ($M_\mathrm{env}\lesssim 10^{-5}M$) and
the mean molecular weight of the stellar material in the envelope does not
depend on the radius $r$.
Thus, the surface mass fractions of hydrogen $X_s$ and helium $Y_s$
given in the table are the same for all mass zones of hydrodynamical
models.

Outer layers of hydrostatically equilibrium LBV--stars are very close
to the boundary of dynamical instability due to the large radiation
pressure, so that approximation errors of finite--difference equations
can lead to expansion of outer mass zones with velocity higher than
the local escape velocity.
In such a case the oscillations can be computed only within a too short
time interval because when the radius of the upper boundary becomes
about several dozen times its initial value the iteration solution
of implicit difference equations do not converge.
To overcome this obstacle one should diminish approximation errors of the
difference equations.
After a number of test computations it was found
that the appropriate solution can be obtained for the number of
mass zones ranging within $10^3\le N\le 3\cdot 10^3$.
In all hydrodynamic models the size of the mass interval decreases to
the inner boundary in order to provide enough approximation at the core
boundary where the gas temperature, pressure and density undergo the
sharp rise to the center.

It is assumed that the radius and luminosity at the inner boundary
remain constant, that is
$\partial r_0/\partial t = \partial L_0/\partial t = 0$.
Determination of the inner radius $r_0$ needs the compromise between
the demands of accuracy and the time step limitations imposed by the
Courant stability criterion.
Choosing the location of the inner boundary and the distribution of
mass zones we tried to satisfy the condition that the integration
time step is
$\Delta t\sim 10^{-5}t_\mathrm{dyn}$, where $t_\mathrm{dyn}=(R^3/GM)^{1/2}$
is the dynamical time scale of the star and $G$ is the
Newtonian constant of gravitation.
Thus, the ratio of the radius of the inner boundary to the
equilibrium radius of photosphere is $0.01\le r_0/R\le 0.05$.
The gas temperature at the inner boundary ranges within
$5\cdot 10^5\K\le T_0\le 10^6\K$, so that
therminuclear energy sources can be ignored.

Hydrodynamic computations with the large number of mass zones
are time consuming but they allowed us to obtain the solution
of the equations of hydrodynamics within time intervals
as long as $10^4$~day
and to apply the discrete Fourier transform for determination
of the mean period $\Pi$ of radial stellar pulsation.
It should be noted however that due to irregular dynamical behaviour
the expansion velocity of the upper boundary sometimes exceeded
the local escape velocity.
In such cases one or a few mass zones excluded from the model and
calculations were continued with the smaller number of mass zones $N$.

\begin{figure}
\centerline{\includegraphics[width=13cm]{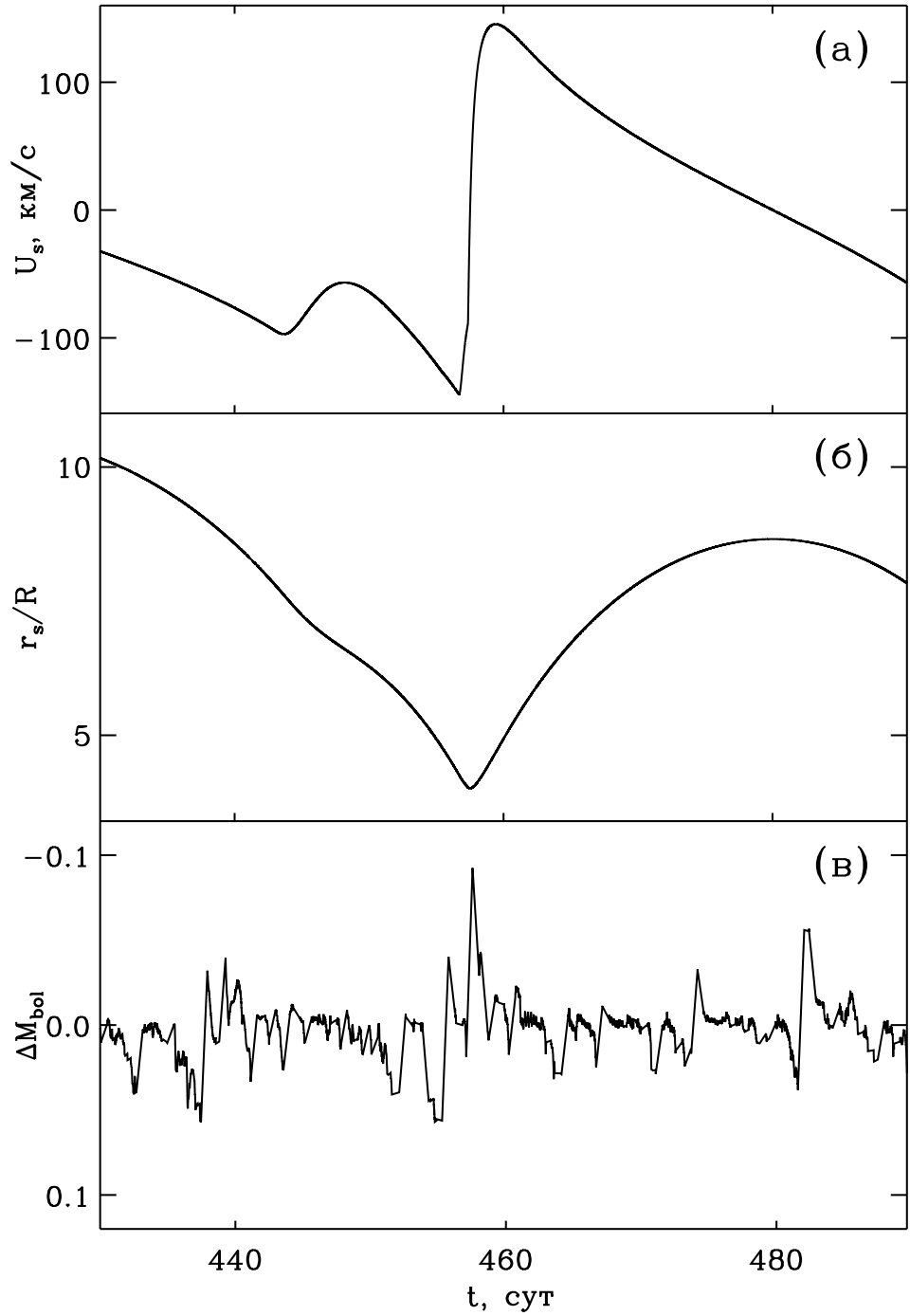}}
\caption{Hydrodynamical model $\mzams=70M_\odot$, $\Teff=3\cdot 10^4\K$
($M=35.7 M_\odot$, $L=1.2\cdot 10^6L_\odot$).
(a) -- Gas flow velocity at the upper boundary of the model $U_\mathrm{s}$;
(b) -- the radius of the upper boundary $r_\mathrm{s}$ in units of
the equilibrium radius of photosphere;
(c) -- bolometric light $\dmbol$.
}
\label{fig2}
\end{figure}

All the models of LBV--stars considered in the present study
were found to be unstable against radial oscillations, the
amplitude growth time being comparable with dynamical time scale
$t_\mathrm{dyn}$.
The amplitude growth ceases at the amplitude $\delta r_\mathrm{s} \sim R$,
so that radial oscillations of LBV--stars are strongly nonlinear.
The nonlinear radial oscillations are illustrated
in Fig.~\ref{fig2}(a) and Fig.~\ref{fig2}(b) where
the temporal dependences of the velocity
$U_\mathrm{s}$ and the radius $r_\mathrm{s}$ of the upper boundary
are shown for the model $\mzams=70M_\odot$, $\Teff=3\cdot 10^4\K$.

Notwithstanding the large amplitude of the radial displacement
the amplitude of light changes is $\le 0.2$~mag.
Here one should note that the rapid light variations
(for the model in Fig.~\ref{fig2} with characteristic time $\lesssim 1$~day)
are due to the discrete nature of the hydrodynamical model
and therefore they are not connected with stellar pulsation.

\begin{figure}
\centerline{\includegraphics[width=15cm]{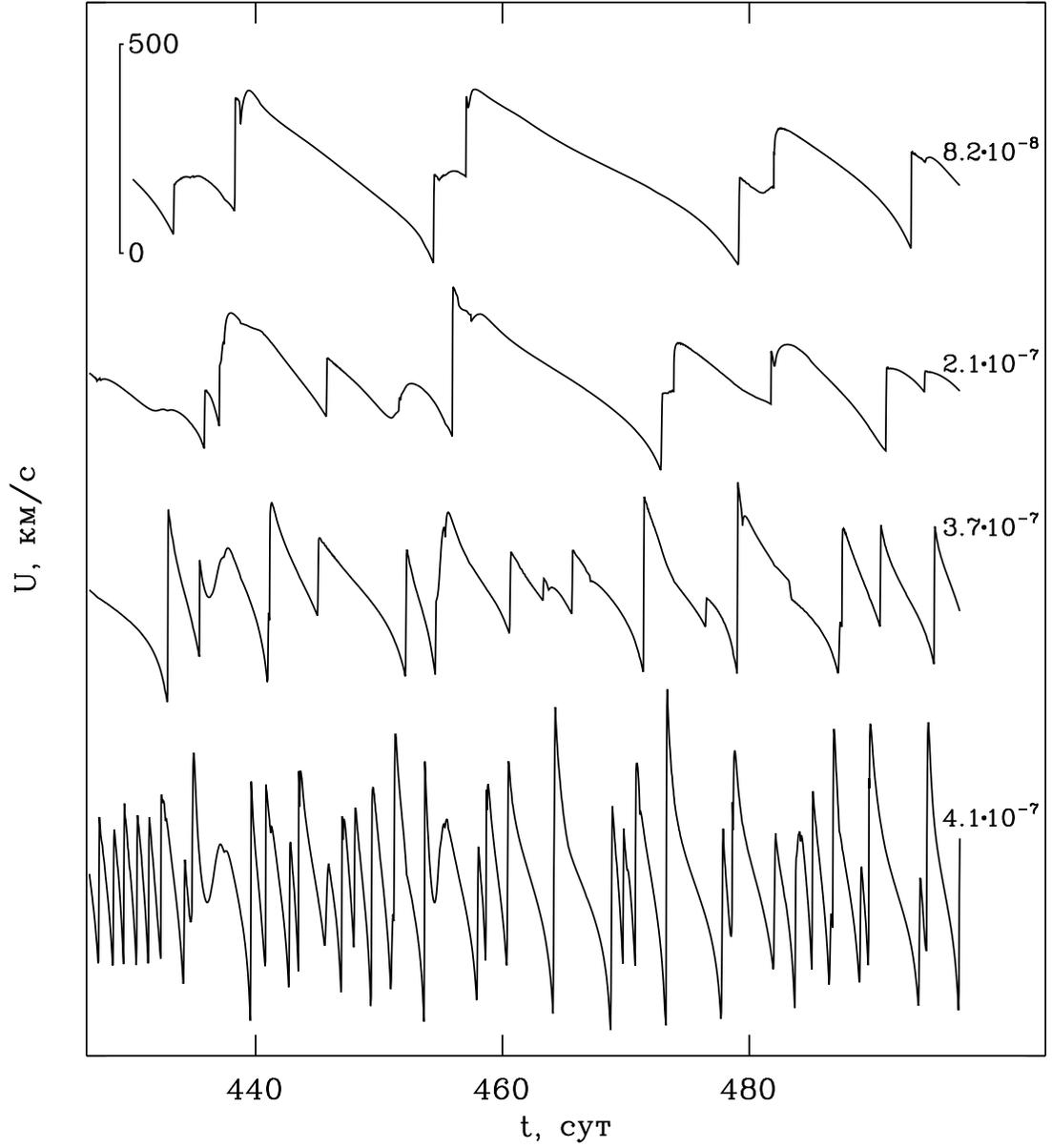}}
\caption{The gas flow velocity $U$ in some mass zones of the
same model as in Fig.~\ref{fig2}.
The plots are arbitrarily shifted along the vertical axis.
Attached at the plots is the Lagrangean coordinate $1 - M_r/M$.
}
\label{fig3}
\end{figure}

In contrast to the most classical radially pulsating stars
where nonlinear effects become important in the outer
layers (e.g., W~Vir and Mira--type variables)
radial oscillations of LBV--stars are nonlinear within the
entire envelope.
This is due to the small pressure gradient and approximately
constant gas density between the core boundary and the
photosphere.
Large amplitude oscillations in different layers of the pulsating
envelope are shown in Fig.~\ref{fig3} where plots of the gas flow
velocity $U$ are given for several mass zones of the hydrodynamical
model.
For the sake of convenience the plots are arbitrarily
shifted along the vertical axis.
The lowest plot in Fig.~\ref{fig3} corresponds to the layers
with the mean radius $r < 0.1R$.

\begin{figure}
\centerline{\includegraphics[width=15cm]{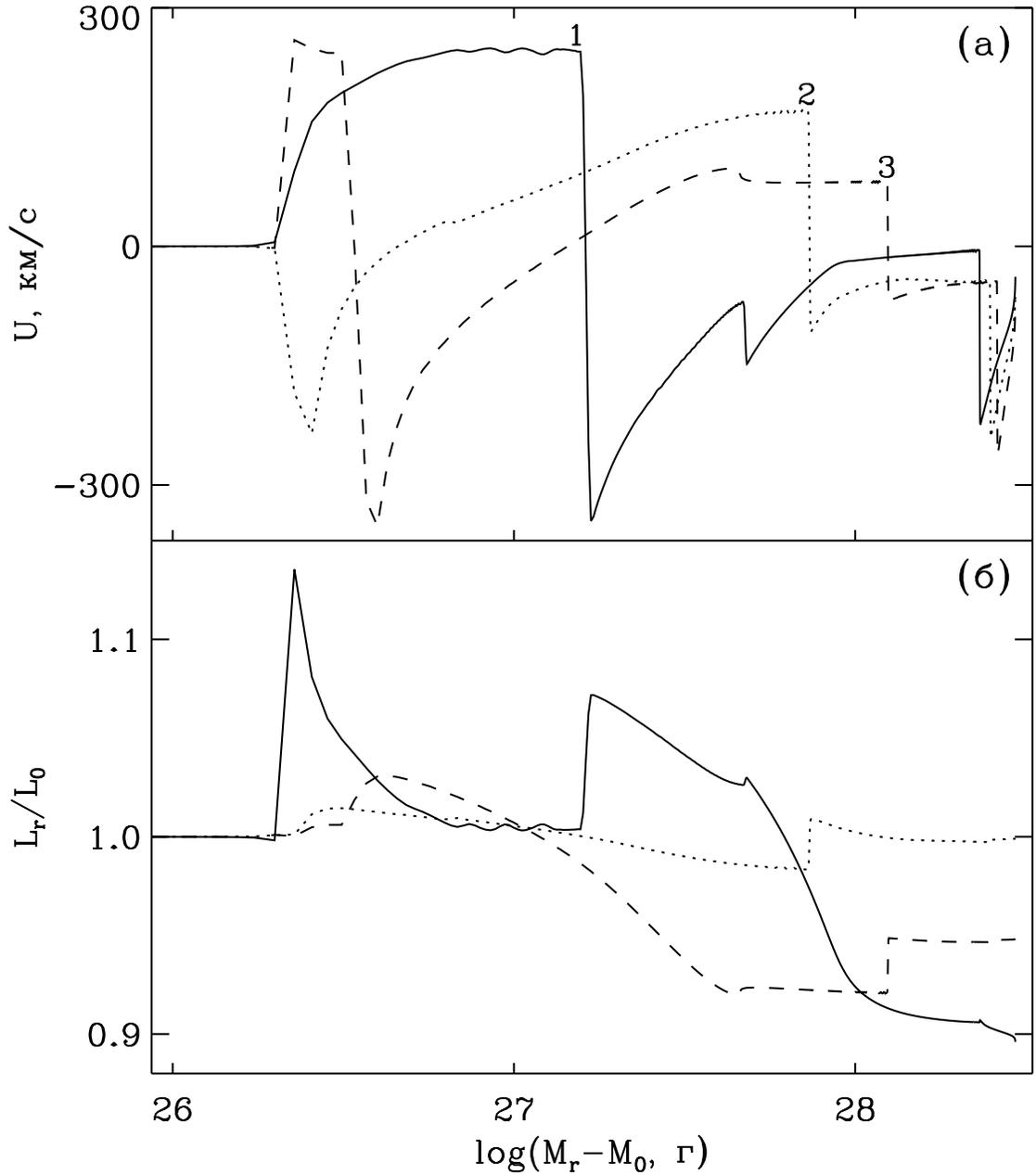}}
\caption{
(a) -- The gas flow velocity $U$;
(b) -- the luminosity $L_r$ in units of the equilibrium luminosity $L_0$
as a function of the Lagrangean coorinate measured from the inner
boundary.
The hydrodynamics model is the same as in Fig~\ref{fig2} and Fig.~\ref{fig3}.
In solid, dotted and slashed lines are shown the plots for
$t_1$, $t_2$ and $t_3$.
}
\label{fig4}
\end{figure}

The main parameter of the radially pulsating star is the pulsation
period $\Pi$ but calculation of this quantity for hydrodynamic
models of LBV--stars is complicated not only due to irregular
oscillations.
The problem is that the most of the mass is
confined inside the compact core and the mass of the
envelope surrounding the core is $M_\mathrm{env}\lesssim 10^{-5}M$,
whereas its extension is about 0.9 of the stellar radius.
Because of the low gas density and the small sound speed
the characteristic motion time of the envelope layers
is significantly longer than that of the core boundary,
so that radial oscillations of LBV--stars are rather the
nonlinear running waves propagating from the inner core to
the stellar surface.
Effects of running waves are illustrated in Fig.~\ref{fig4}(a),
where three plots of the gas flow velocity $U$ are shown
as a function of the Lagrangean coordinate measured from
the inner boundary.
The expansion of inner layers occurs approximately at
$t=t_1$ and $t=t_2$ but displacement of the running wave
during the time interval $t_2 - t_1$
is significantly less than the stellar radius.
Thus, the running wave reaches the upper boundary during several
oscillation periods of inner layers.
Dependence of the oscillation period on the spatial coordinate
is clearly seen from velocity plots corresponding to
different depths inside the envelope (Fig.~\ref{fig3}).

\begin{figure}
\centerline{\includegraphics[width=15cm]{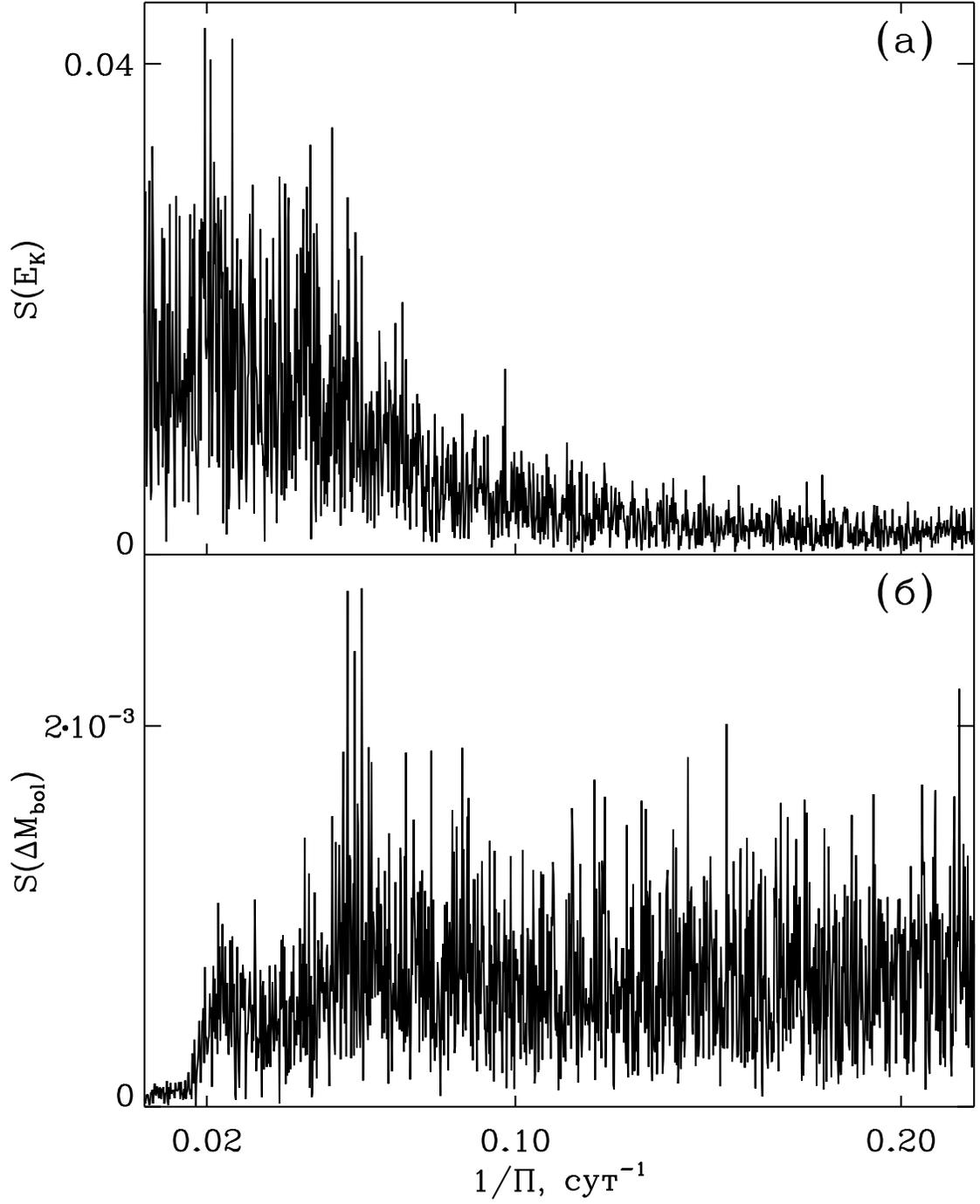}}
\caption{The hydrodynamical model
$\mzams=80M_\odot$, $\Teff=2\cdot 10^4\K$
($M=42.3M_\odot$, $L=1.6\cdot 10^6L_\odot$).
(a) -- The Fourier spectrum of the kinetic energy of pulsating
envelope $S(\EK)$;
(b) -- the Fourier spectrum of the bolometric light $S(\dmbol)$.
}
\label{fig5}
\end{figure}

The bolometric radiation flux from the upper boundary
depends on the contribution of layers at different depths
with different oscillation periods.
That is why light variations rather weakly correlate with variations
of the upper boundary velocity (see Fig.~\ref{fig2}).
The pulsation period calculated from the discrete Fourier transform
of the kinetic energy $\EK$ of the pulsating envelope is always
longer than that evaluated from the light curve.
This is due to the fact that the most of the kinetic energy
is contributed by outer layers, whereas the light variations
depend on deeper layers with shorter periods.
The typical Fourier spectra of the kinetic energy $S(\EK)$ and
bolometric light $S(\Delta M_\mathrm{bol})$ are shown in Fig.~\ref{fig5}
for the model $\mzams=80M_\odot$, $\Teff=2\cdot 10^4\K$.
The mean period of variations of the kinetic energy is $\Pi(\EK)=51.2$~day
while the mean light period is $\Pi=16.6$~day.

Last two columns of the table give the mean periods of light changes
$\Pi$ and corresponding pulsation constants $Q$.
Here one should bear in mind that in contrast to radial pulsations
in the form of standing waves
the pulsation constants of LBV--stars cannot be considered as
their mechanical characteristics.

The sharp increase of the luminosity $L_r$ at the front of the running wave
propagating to the upper boundary (see Fig.~\ref{fig4}) is the principal
cause of the change of radiation flux emerging from the upper boundary.
Thus, the light variations of LBV--stars are due to dissipation of the
kinetic energy of running waves rather than due to $\kappa$--mechanism.
Indeed, modulation of the radiation flux by $\kappa$--mechanism can
arise in the vicinity of Z--bump ($T\approx 2\cdot 10^5\K$)
but in LBV--stars these layers are at the boundary of the compact core,
so that their mass
is too small for driving the pulsational instability.
For some models we carried out hydrodynamical calculations with different
location of the inner boundary and found that models with temperature
at the inner boundary $T\sim 10^5\K$ demonstrate instability similar
to that obtained for models with deeper inner boundary locating below
the Z--bump.
Therefore, the pulsational instability is due to the fact that
the adiabatic exponent within the envelope is close to the
critical value $\Gamma_1=4/3$.
The inefficiency of the $\kappa$--mechanism in LBV--stars was
considered in the framework of the linear theory by Kiriakidis et al. (1993).

\section*{conclusion}

In the present study we computed the hydrodynamical models of
radially oscillating LBV--stars  with effective temperatures
$1.5\cdot 10^4\K\le\Teff\le 3\cdot 10^4\K$ and this allows us
to conclude that micro\-variability is most probably due to
radial oscillations.
The $\kappa$--mechanism is not responsible for pulsational
instability and oscillations appear is due thermodynamical properties
of the stellar material with adiabatic exponent close to its
critical value $\Gamma_1 = 4/3$.
Strongly nonadiabatic oscillations prevent the development of the
dynamical instability due to large radiative losses accompanying
the motion of the gas.

Nonlinear radial oscillations of LBV--stars exist in the form of
nonlinear running waves propagating from the boundary of the
compact core to the stellar surface, the amplitude of radial velocity
variations being as large as several hundred km/s.
The amplitude of theoretical bolometric light curve is
($\Delta M_\mathrm{bol}\le 0.2$~mag) seems to be overestimated due to
limited spatial resolution of hydrodynamical models.

Mean periods estimated from hydrodynamical models
($6~\mathrm{day}\le\Pi\le 31~\mathrm{day}$)
are in agreement with observations.
For example, observational estimates of microvarion periods are as follows:
$\Pi(\mathrm{P~Cyg})\approx 17.3$~day (de Groot et al., 2001),
$\Pi(\mathrm{B416~M33})\approx 8.26$~day (Shemmer et al., 2000),
$20~\mathrm{day}\le\Pi(\mathrm{v532~M31})\le 30~\mathrm{day}$ (Sholukhova et al., 2002),
$25.7~\mathrm{day}\le\Pi(\zeta^1~\mathrm{Sco})\le 32~\mathrm{day}$ (Sterken et al., 1997).
The only exception is $\eta$~Car
but its too long period of $\Pi=58.58$~day (Sterken et al., 1996)
is rather due to extremely high luminosity ($L\sim 5\cdot 10^6L_\odot$)
of this star.

\newpage
\section*{References}

\begin{enumerate}

\item G.A. Vazquez, C. Leitherer, et al., Astrophys. J., 663, 995 (2007).

\item J.S. Vink, A. de Koter, IAUS, \textbf{212}, 259 (2003).

\item J.S. Vink, A. de Koter, H.J.G.L.M. Lamers, Astron.Astrophys., 362, 295 (2000)

\item J.S. Vink, A. de Koter, H.J.G.L.M. Lamers, Astron.Astrophys., 369, 574 (2001)

\item A. Gal--Yam, D.C. Leonard, D.B. Fox, et al., Astrophys. J., \textbf{656}, 372 (2007).

\item A.M. van Genderen, Astron.Astrophys., \textbf{208}, 135 (1989)

\item A.M. van Genderen, Astron. Astrophys., \textbf{366}, 508 (2001).

\item A.M. van Genderen, P.S. Th\'e, Space Sci. Rev., \textbf{39}, 317 (1984).

\item W. Glatzel, M. Kiriakidis, MNRAS, \textbf{263}, 375 (1993).

\item M. de Groot, IrAJ, \textbf{18}, 163 (1988).

\item M. de Groot, C. Sterken, A.M. van Genderen, Astron. Astrophys., \textbf{376}, 224 (2001).

\item W. Dziembowski, J. Slawinska, Acta Astron., \textbf{55}, 195 (2005).

\item E.A. Dorfi, M.U. Feuchtinger, A. Gautschy, ASP Conf. Ser., \textbf{203}, 109 (2000).

\item E.A. Dorfi, A. Gautschy, Comm. in Asteroseismology, \textbf{141}, 57 (2002).

\item M. Kiriakidis, K.J. Fricke, W. Glatzel, MNRAS, \textbf{264}, 50 (1993).

\item Lamers, H.J.G.L.M. 1987, in \textsl{Instabilities in Luminous Early--Type Stars},
      ed. H.J.G.L.M. Lamers, and C.W.H. de Loore (Dordrecht: Reidel), p. 99.

\item N. Langer, W.-R. Hamann, M. Lennon, et al., Astron. Astrophys., \textbf{290}, 819 (1994).

\item C. Leitherer, N.A. Damineli, W. Schmutz, ASP Conf. Ser., \textbf{22}, 366 (1992)

\item A. Maeder, Astron. Astrophys., \textbf{120}, 113 (1983).

\item N. Nugis, H. J. G. L. M. Lamers, Astron. Astrophys., \textbf{360}, 227 (2000).

\item H. Nieuwenhuijzen and C. de Jager, Astron.Astrophys., \textbf{231}, 134 (1990).

\item N., Smith, M. Ganeshalingam, R. Chornock, et al., Astrophys. J., \textbf{697}, L49 (2009).

\item C. Sterken, M.J.H. de Groot, A.M. van Genderen, Astron. Astrophys. Suppl. Ser., \textbf{116}, 9 (1996).

\item C. Sterken, M.J.H. de Groot, A.M. van Genderen, Astron. Astrophys., \textbf{326}, 640 (1997).

\item C. Trundle, R. Kotak, J.S. Vink, et al., Astron. Astrophys., \textbf{483}, L47 (2008).

\item Yu.A. Fadeyev, Ast.L., \textbf{33}, 645 (2007).

\item Yu.A. Fadeyev, Ast.Rep., \textbf{52}, 645 (2008a).

\item Yu.A. Fadeyev, Ast.L., \textbf{34}, 772 (2008b).

\item D.J. Frew, J. Astron. Data, \textbf{10}, 6 (2004).

\item R.M. Humphreys, K. Davidson, PASP, \textbf{106}, 1025 (1994).

\item O. Shemmer, E.M. Leibowitz, P. Szkody, MNRAS, \textbf{311}, 698 (2000).

\item O. Sholukhova, A. Zharova, S. Fabrika, D. Malinovskii, ASP Conf. Ser., \textbf{259}, 522 (2002).

\item C. de Jager, Astron. Astrophys. Rev., \textbf{8}, 145 (1998).
\end{enumerate}

\newpage
\begin{table}
\caption{Models of LBV--stars.}
\begin{tabular}{r|r|r|r|r|r|r|r|r}
\hline
$\mzams/M_\odot$ & $\Teff, 10^3\K$ & $L/L_\odot$ & $M/M_\odot$ & $R/R_\odot$ & $\xs$ & $\ys$ & $\Pi$, day & $Q$, day\\
\hline
 70  &  15  &  $1.24\cdot 10^6$  &  36.02  & 165 &  0.23  &  0.75  &  28  &  0.079 \\  % 165.2
     &  20  &  $1.24\cdot 10^6$  &  35.82  &  93 &  0.21  &  0.77  &  16  &  0.107 \\  % 93.15
     &  30  &  $1.24\cdot 10^6$  &  35.68  &  41 &  0.16  &  0.82  &   6  &  0.135 \\  % 41.34
 80  &  15  &  $1.63\cdot 10^6$  &  42.48  & 190 &  0.12  &  0.86  &  30  &  0.074 \\  % 190.4
     &  20  &  $1.60\cdot 10^6$  &  42.33  & 105 &  0.12  &  0.86  &  17  &  0.102 \\  % 105.4
     &  25  &  $1.56\cdot 10^6$  &  42.13  &  67 &  0.12  &  0.86  &  10  &  0.119 \\  % 66.71
     &  30  &  $1.51\cdot 10^6$  &  41.84  &  46 &  0.11  &  0.87  &   6  &  0.126 \\  % 45/71
 90  &  15  &  $1.90\cdot 10^6$  &  49.06  & 205 &  0.11  &  0.87  &  31  &  0.074 \\  % 204.7
     &  20  &  $1.86\cdot 10^6$  &  48.80  & 105 &  0.11  &  0.87  &  16  &  0.092 \\  % 114.2
     &  25  &  $1.80\cdot 10^6$  &  47.81  &  72 &  0.11  &  0.87  &  12  &  0.137 \\  % 71.73
     &  30  &  $1.78\cdot 10^6$  &  47.18  &  50 &  0.09  &  0.89  &   7  &  0.137 \\  % 49.87
\hline
\end{tabular}
\end{table}
\clearpage

\newpage

\newpage

\newpage

\newpage

\newpage

\end{document}